\documentstyle[12pt,procsla,epsfig]{article}
  
\catcode`\@=11
\long\def\@makefntext#1{
\protect\noindent \hbox to 3.2pt {\hskip-.9pt
$^{{\ninerm\@thefnmark}}$\hfil}#1\hfill}		

\def\@makefnmark{\hbox to 0pt{$^{\@thefnmark}$\hss}}  
	
\def\ps@myheadings{\let\@mkboth\@gobbletwo
\def\@oddhead{\hbox{}
\rightmark\hfil\ninerm\thepage}   
\def\@oddfoot{}\def\@evenhead{\ninerm\thepage\hfil
\leftmark\hbox{}}\def\@evenfoot{}
\def\sectionmark##1{}\def\subsectionmark##1{}}
\def \lta {\mathrel{\vcenter
{\hbox{$<$}\nointerlineskip\hbox{$\sim$}}}} \def \gta
{\mathrel{\vcenter {\hbox{$>$}\nointerlineskip\hbox{$\sim$}}}}
\begin{document}

\centerline{\normalsize\bf BREAKTHROUGHS ON THE DARK MATTER ISSUE}
\baselineskip=16pt
\vspace*{0.6cm}
\centerline{\footnotesize ANTONIO MASIERO and FRANCESCA ROSATI}
\baselineskip=13pt
\centerline{\footnotesize\it SISSA, Via Beirut 2-4, I-34013 Trieste, ITALY}
\centerline{\footnotesize\it INFN, sezione di Trieste, Padriciano 99, 
            I-34012 Trieste, ITALY}
\baselineskip=12pt
\centerline{\footnotesize E-mail: masiero@sissa.it, rosati@sissa.it}
\vspace*{0.9cm}
\abstracts{
Last year observations had a profound impact on our views on the amount 
and nature of dark matter in the universe.
We give a brief review of the recent history of dark matter models 
beyond the pure cold dark matter universe. 
In view of the most recent cosmological data, we then go on to discuss 
models with a positive cosmological constant. 
Finally we explicitly analyse a class of particle physics models for a 
dynamical cosmological component with negative pressure (``quintessence''), 
in the context of supersymmetric theories.}

\normalsize\baselineskip=15pt
\setcounter{footnote}{0}
\renewcommand{\thefootnote}{\alph{footnote}}

\section{Introduction}

\subsection{The Universe with $\Lambda$}

The year 1998 has witnessed major changes in our perspectives on the
amount and nature of the different components of dark matter (DM)
in the universe.

The indications for a presently accelerating universe coming from 
redshift-distance measurements of High-Z Supernovae Ia 
(SNe Ia)\cite{scp,highz}, combined with cosmic microwave background 
(CMB) data\cite{cmb} and cluster mass distribution\cite{cluster} seem 
to favour models with a non-vanishing cosmological constant 
$\Lambda$. 
Indeed the energy density contributed by $\Lambda$ should roughly be
twice as much as the enegy density of the matter of the whole universe, 
thus leading to a flat universe (with $\Omega_{total} = 1$) composed by
$1/3$ of matter and the remaining $2/3$ of cosmological constant or
vacuum energy.

If these results have greatly excited the community of astrophysicists 
and cosmologists, they have not left indifferent the particle physicists.
Indeed, most of the people belonging to latter group used to think that 
the famous problem of the smallness of the cosmological constant would 
have eventually found a solution with the discovery of some symmetry that 
could make it vanishing. 
On the contrary, the 1998 observational evidence seems to indicate that
$\Lambda$, far from being zero, could instead constitute the major source 
of energy of a flat (critical) universe. 

But what is actually $\Lambda$? 
Last year data support the view that the  universe is presently dominated 
by a smooth component with effective negative pressure; this is in fact the 
most general requirement in order to explain the observed accelerated expansion. 
The most straightforward candidate for that is, of course, a ``true'' 
cosmological constant\cite{carroll2}, but a plausible alternative that has 
recently received a great deal of attention is a dynamical vacuum energy 
given by a scalar field rolling down its potential.
A cosmological scalar field, depending on its dynamics, can easily fulfill 
the condition of an equation of state $w =p/ \rho$ between
$-1$ (which corresponds to the cosmological constant case) and $0$
(that is the equation of state of matter). 
Since it is useful to have a short name for the rather long definition 
of this dynamical vacuum energy, we follow the literature in calling it 
briefly ``quintessence''\cite{quint1}.  

The major bulk of this talk will be devoted to the discussion of  the
``quintessence'' possibility in the context of supersymmetric (SUSY) theories. 
But before plunging into this, let us remind the readers that the 
abovementioned data should be taken with some caution.

\subsection{The alternatives}

It is indeniably impressive that having
three sets of independent data depending on two parameters -- the fraction
densities in matter and cosmological constant, $\Omega_m$ and
$\Omega_{\Lambda}$ -- we get an overconstrained system with a unique solution 
(what has been called with ironical emphasis ``cosmic concordance''\cite{turner4}). 
However, one should not underestimate the fact that for
some of the measurements -- for instance the position of the first 
acoustic peak in the CMB and the study of the systematics in the SNe Ia -- 
we are still at a preliminary level. 
While waiting for further observational evidence in favour of cosmologies based 
on cold DM (CDM) and quintessence (also called QCDM models), one should then 
anyway keep an open eye on the alternative solutions to the dark matter problem 
which have emerged after the crisis of the pure CDM standard model.

In the pure CDM model, almost  all of the energy density needed to reach the 
critical one (the remaining few percent being given by 
the baryons) was provided by cold dark matter alone. 
However, some observational facts (in particular the results of COBE)
put this model into trouble, showing that it cannot correctly reproduce the 
power spectrum of density perturbations at all scales. 
At the same time it became clear that some amount of CDM was needed anyway 
in order to obtain a successful scheme for large scale structure formation.

A popular option is that of a flat universe realized with the total energy 
density mostly provided by two different matter components, CDM and hot DM (HDM) 
in a convenient fraction. 
These models, which have been called mixed DM (MDM)\cite{MDM}, succeeded to fit 
the entire power spectrum quite well, although -- having $\Omega_{m}= 
\Omega_{tot}=1$ -- they obviously can't account for the most recent data.    

Another interesting possibility for improving CDM models consists in the 
introduction of some late time decaying particle\cite{kim}.
The injection of non-thermal radiation due to such decays and the
consequent increase of the horizon length at the equivalence time 
could lead to a convenient suppression of the excessive power at small scales
(hence curing the major desease of the pure CDM standard model). 
As appealing as this proposal may be from the cosmological point of view, its
concrete realization in particle physics models meets several
difficulties. Indeed, after considering cosmological and astrophysical
bounds on such late decays, it turns out that only few candidates survive
as viable solutions (for a recent analysis in the context of SUSY
extensions of the SM with or without R parity see Ref.\cite{marco}).

Another route which has been followed in the attempt to go beyond the 
pure CDM proposal is the possibility of having some form of warm DM (WDM). 
The implementation of this idea is quite attractive in SUSY models where the
breaking of SUSY is conveyed by gauge interactions instead of gravity
(these are the so-called gauge mediated SUSY breaking (GMSB) models). 
In these schemes the gravitino mass loses its role of fixing the typical 
size of soft breaking terms and we expect it to be much smaller than in 
the more traditional supergravity models. 
The gravitino in GMSB theories can behave as a WDM candidate. 
Unfortunately, critical universes with pure WDM are known to
suffer from serious troubles\cite{colombi}. It has been shown that even 
variants of the pure WDM models (with an additional HDM component or with 
a non-vanishing cosmological constant) still exhibit various problems in 
correctly reproducing the large scale structure data\cite{borgani}. 

Finally, the alternative to pure CDM which succeeded to fit the whole power 
spectrum as well as the MDM models was provided by the so-called $x$CDM 
models\cite{turner1}.
In those schemes, CDM is accompanied with a cosmological constant-like 
contribution (named $x$) adding up to reach the critical energy density. 
In view of the recent observations that we presented above, it is clear that 
$x$CDM models today enjoy the major success. We now turn to discuss these schemes 
in the context of SUSY theories\cite{noi}.

\section{Supersymmetry and Quintessence}

As already said in the introduction, the 1998 indications for
an accelerating universe have recently drawn a great deal of attention on
cosmological models with $\Omega _{m}\sim 1/3$ and $\Omega _{\Lambda}\sim 2/3$.
More generally, the r\^ole of the cosmological constant in accelerating the
universe expansion could be played by any smooth component with
negative equation of state $p_{Q}/\rho _{Q}=w_{Q}\lta -0.6$\cite{quint1,friem}, 
as in the so-called ``quintessence'' models
(QCDM)\cite{quint1}, otherwise known as $x$CDM models\cite{turner1}. 

\subsection{Scalar field cosmology}

A natural candidate for quintessence is given by a rolling scalar
field $Q$ with potential $V(Q)$ and equation of state
\[
w_{Q}=\frac{ \dot{Q}^{2}/2 -V(Q)}{ \dot{Q}^{2}/2 +V(Q)}\;,
\]
which -- depending on the amount of kinetic energy -- could in
principle take any value from $-1$ to $+1$.  The study of scalar field
cosmologies has shown\cite{rp,liddle} that for certain potentials
there exist attractor solutions that can be of the ``scaling''
\cite{wet,cop,fj} or ``tracker''\cite{zws,swz} type; that means that
for a wide range of initial conditions the scalar field will rapidly
join a well defined late time behavior. 

If $\rho _{Q}\ll \rho _{B}$, where $\rho _{B}$ is the energy
density of the dominant background (radiation or matter), the
attractor can be studied analytically. 

In the case of an exponential potential, $V(Q)\sim \exp {(-Q)}$ the
solution $Q\sim \ln {t}$ is, under very general conditions, a ``scaling''
attractor in phase space characterized by $\rho _{Q}/\rho _{B}\sim
{\rm const}$\cite{wet,cop,fj}. 
This could potentially solve the so called ``cosmic coincidence'' problem, 
providing a dynamical explanation for the order of magnitude equality 
between matter and scalar field energy today.  
Unfortunately, the equation of state for this attractor is $w_{Q}=w_{B}$, 
which cannot explain the acceleration of the universe neither during RD 
($ w_{rad}=1/3$) nor during MD ($w_m=0$).  
Moreover, Big Bang nucleosynthesis constrain the field
energy density to values much smaller than the required $ \sim 2/3$
\cite{liddle,cop,fj}. 

If instead an inverse power-law potential is considered,
$V(Q)=M^{4+\alpha }Q^{-\alpha }$, with $\alpha >0$, the attractor
solution is $Q\sim t^{1-n/m}$, where $n=3(w_{Q}+1)$, $m=3(w_{B}+1)$;
and the equation of state turns out to be $w_{Q}=(w_{B}
\,\alpha-2)/(\alpha+2)$, which is always negative during MD. 
The ratio of the energies is no longer constant but scales as\ 
$\rho _{Q}/\rho_{B}\sim a^{m-n}$ thus growing during the cosmological 
evolution, since $n$ $<m$.  
$\rho _{Q}$ could then have been safely small during
nucleosynthesis and have grown lately up to the phenomenologically
interesting values.These solutions are then good candidates for
quintessence and have been denominated ``tracker'' in the literature
\cite{liddle,zws,swz}. 

The inverse power-law potential does not improve the cosmic
coincidence problem with respect to the cosmological constant
case. Indeed, the scale $M $ has to be fixed from the requirement that
the scalar energy density today is exactly what is needed. This
corresponds to choosing the desired tracker path.  An important
difference exists in this case though.  
The initial conditions for the physical variable $\rho _{Q}$ can vary 
between the present critical energy density $\rho_{cr}^0$ and the 
background energy density $\rho_B$ at the time of beginning\cite{swz} 
(this range can span many tens of orders of magnitude, depending on the 
initial time), and will anyway end on the tracker path before the present 
epoch, due to the presence of an attractor in phase space\cite{zws,swz}.  
On the contrary, in the cosmological constant case, the physical variable
$\rho _{\Lambda }$ is fixed once for all at the beginning. This allows
us to say that in the quintessence case the fine-tuning issue, even if
still far from solved, is at least weakened. 

A great effort has recently been devoted to find ways to constrain
such models with present and future cosmological data in order to
distinguish quintessence from $\Lambda $ models\cite{constr,det}.  
An even more ambitious goal is the partial reconstruction of the scalar field
potential from measuring the variation of the equation of state with
increasing redshift\cite{turner2}. 

On the other hand, the investigation of quintessence models from the
particle physics point of view is just in a preliminary stage and a
realistic model is still missing (see for example Refs. 
\cite{bin,pngb,lyth,sugra}). 
There are two classes of problems; the construction of a field theory model 
with the required scalar potential and the interaction of the quintessence 
field with the standard model (SM) fields\cite{car}.  
The former problem was already considered by Bin\'{e}truy\cite{bin}, 
who pointed out that scalar inverse power law potentials appear in 
supersymmetric QCD theories (SQCD)\cite{SQCD} with $N_{c}$ colors 
and $N_{f}<N_{c}$ flavors. 
The latter seems the toughest. Indeed the quintessence field today has 
typically a mass of order $H_{0}\sim 10^{-33}$eV. 
Then, in general, it would mediate long range interactions of gravitational 
strength, which are phenomenologically unacceptable. 

In the remaining part of the talk, we will address in more details these 
problems in the framework of SQCD, following the work done in Ref.\cite{noi}.

\subsection{Susy QCD}

As already noted by Bin\`{e}truy\cite{bin}, supersymmetric QCD
theories with $N_{c}$ colors and $N_{f}<N_{c}$ flavors\cite{SQCD} may
give an explicit realization of a model for quintessence with an
inverse power law scalar potential. 
The remarkable feature of these theories is that the superpotential is 
exactly known non-perturbatively. Moreover, in the range of field values 
that will be relevant for our purposes (see below) quantum corrections to the
K\"{a}hler potential are under control. 
As a consequence, we can study the scalar potential and the field equations 
of motion of the full quantum theory, without limiting ourselves to the 
classical approximation. 

The matter content of the theory is given by the chiral superfields
$Q_{i}$ and $\overline{Q}_{i}$ ($i=1\ldots N_{f}$) transforming
according to the $ N_{c}$ and $\overline{N}_{c}$ representations of
$SU(N_c)$, respectively.  In the following, the same symbols will be
used for the superfields $Q_{i}$, $\overline{Q}_{i}$, and their scalar
components. 

Supersymmetry and anomaly-free global symmetries constrain the
superpotential to the unique {\it exact} form
\begin{equation}
W=(N_{c}-N_{f})\left( \frac{\Lambda ^{(3N_{c}-N_{f})}}{{\rm
det}T}\right) ^{ \frac{1}{N_{c}-N_{f}}} \label{superpot}
\end{equation}
where the gauge-invariant matrix superfield $T_{ij}=Q_{i}\cdot
\overline{Q}_{j}$ appears. $\Lambda $ is the only mass scale of the
theory.  
It is the supersymmetric analogue of $\Lambda _{QCD}$, the
renormalization group invariant scale at which the gauge coupling of
$SU(N_{c})$ becomes non-perturbative. As long as scalar field values
$Q_{i},\overline{Q}_{i}\gg $ $\Lambda $ are considered, the theory is
in the weak coupling regime and the canonical form for the K\"{a}hler
potential may be assumed.  
The scalar and fermion matter fields have then canonical kinetic terms, 
and the scalar potential is given by
\begin{equation}
V(Q_{i},\overline{Q}_{i})=\sum_{i=1}^{N_{f}}\left(
|F_{Q_{i}}|^{2}+|F_{\overline{Q}_{i}}|^{2}\right)
+\frac{1}{2}D^{a}D^{a} \label{potscal}
\end{equation}
\label{scalarpot} 
where $F_{Q_{i}}=\partial W/\partial Q_{i}$, 
$F_{\overline{Q}_{i}}=\partial W/\partial \overline{Q}_{i}$, and
\begin{equation}
D^{a}=Q_{i}^{\dagger }t^{a}Q_{i}-\overline{Q}_{i}t^{a}\overline{Q}
_{i}^{\dagger }\;.  \label{d-terms}
\end{equation}
The relevant dynamics of the field expectation values takes place
along directions in field space in which the above D-term vanish, {\it
i.e.} the perturbatively flat directions $\langle Q_{i\alpha }\rangle
=\langle \overline{Q}_{i\alpha }^{\dagger }\rangle $, where $\alpha
=1\cdots N_{c}$ is the gauge index.  
At the non-perturbative level these directions get a non vanishing 
potential from the F-terms in (\ref{potscal}), which are zero at any order 
in perturbation theory.  
Gauge and flavor rotations can be used to diagonalize the
$\langle Q_{i\alpha }\rangle $ and put them in the form
\[
\langle Q_{i\alpha }\rangle =\langle \overline{Q}_{i\alpha }^{\dagger
}\rangle =
\begin{array}{l}
q_{i}\delta _{i\alpha }\;\;\;\;1\leq \alpha \leq N_{f} \\ 0\;\;\;\;\ \
\ \ \;N_{f}\leq \alpha \leq N_{c}
\end{array}
. 
\]
Along these directions, the scalar potential is given by
\begin{eqnarray*}
v(q_{i}) &\equiv &\langle V(Q_{i},\overline{Q}_{i})\rangle
=2\;\frac{\Lambda ^{2a }}{\prod_{i=1}^{N_{f}}|q_{i}|^{4d
}}\;\left( \sum_{j=1}^{N_{f}}\frac{1}{|q_{j}|^{2}}\right) ,\qquad \\
&& \\ \qquad a &=&\frac{3N_{c}-N_{f}}{N_{c}-N_{f}},\ \ \ \ \ \
d =\frac{1}{N_{c}-N_{f}}. 
\end{eqnarray*}
In the following, we will be interested in the cosmological evolution
of the $N_{f}$ expectation values $q_{i}$, given by
\[
\langle \ddot{Q_{i}}+3H\dot{Q_{i}}+\frac{\partial V}{\partial
Q_{i}^{\dagger }}\rangle =0\;\;,\;i=1,...,N_{f}\;. 
\]
In Ref.\cite{bin} the same initial conditions for all the $N_{f}$
VEV's and their time derivatives were chosen. With this very peculiar
choice the evolution of the system may be described by a single VEV
$q$ (which we take real) with equation of motion 
\begin{equation}
\ddot{q}+3H\dot{q}-g \frac{\Lambda ^{2a }}{q^{2g
+1}}=0\ ,\qquad \qquad g =\frac{N_{c}+N_{f}}{N_{c}-N_{f}}\ ,
\label{onescalar}
\end{equation} 
thus reproducing exactly the case of a single scalar field $\Phi
$ in the potential $V=\Lambda ^{4+2g}\Phi ^{-2g}/2$
considered in Refs.\cite{rp,liddle,swz}.  In this paper we will
consider the more general case in which different initial conditions
are assigned to different VEV's, and the system is described by
$N_{f}$ coupled differential equations. Taking for illustration the
case $N_{f}=2$, we will have to solve the equations

\[
\ddot{q}_{1}+3H\dot{q}_{1}-d\cdot\! \ q_{1}\ \frac{\Lambda ^{2a
}}{\left( q_{1}q_{2}\right) ^{2d N_{c}}}\ \left[
2+N_{c}\frac{q_{2}^{2}}{q_{1}^{2}}\right] =0\ ,
\]
 
\begin{equation}
\ddot{q}_{2}+3H\dot{q}_{2}-d\cdot\! \ q_{2}\ \frac{\Lambda ^{2a
}}{\left( q_{1}q_{2}\right) ^{2d N_{c}}}\ \left[
2+N_{c}\frac{q_{1}^{2}}{q_{2}^{2}}\right] =0\ , \label{eom}
\end{equation}
with $H^{2}=8\pi /3M_P^{2}\ (\rho _{m}+\rho _{r}+\rho _{Q})$, where
$M_P$ is the Planck mass, $\rho _{m(r)}$ is the matter (radiation)
energy density, and $\rho
_{Q}=2(\dot{q}_{1}^{2}+\dot{q}_{2}^{2})+v(q_{1},q_{2})$ is the total
field energy. 

\subsection{The tracker solution}

In analogy with the one-scalar case, we look for power-law{\it \
}solutions of the form
\begin{equation}
q_{tr,i}=C_{i}\cdot t^{\, p_{i}}\ ,\qquad \qquad i=1,\cdots ,\ N_{f}\ . 
\label{scaling}
\end{equation}
It is straightforward to verify that -- when $\rho _{Q}\ll \rho _{B}$
-- the only solution of this type is given by
\[
p_{i}\equiv p=\frac{1-r}{2}\ ,\ \ \ \ \ \ C_{i}\equiv C=\left[
X^{1-r}\ \Lambda ^{2(3-r)}\right] ^{1/4}\ ,\ \ \
\ \ \ i=1,\cdots ,\ N_{f}\ ,
\]
with
\[
X\equiv \frac{4\ m\ \ (1+r)}{(1-r)^{2}\ [12-m(1+r)]}\ ,
\]
where we have defined $r\equiv N_{f}/N_c$ $(=1/N_{c},\ldots ,1-1/N_{c})$.  
This solution is characterized by an equation of state
\begin{equation}
w_{Q}=\frac{1+r}{2}w_{B}-\frac{1-r}{2}\ .  \label{eosfree}
\end{equation}
Eq. (\ref{eosfree}) can be derived as usual from energy conservation
{\it i.e. }$d\ (a^3 \rho_{Q})=-3\ a^2 p_{Q}$. 

Following the same methods employed in Ref.\cite{liddle} one can show
that the above solution is the unique stable attractor in the space of
solutions of eqs. (\ref{eom}). Then, even if the $q_{i}$'s start with
different initial conditions, there is a region in field configuration
space such that the system evolves towards the equal fields solutions
(\ref{scaling}), and the late-time behavior is indistinguishable from
the case considered in Ref.\cite{bin}. 

The field energy density grows with respect to the matter energy
density as
\begin{equation}
\frac{\rho _{Q}}{\rho _{m}}\sim a^{\frac{3(1+r)}{2}},
\end{equation}
where $a$ is the scale factor of the universe. The scalar field energy
will then eventually dominate and the approximations leading to the
scaling solution (\ref{scaling}) will drop, so that a numerical
treatment of the field equations is mandatory in order to describe the
phenomenologically relevant late-time behavior. 

The scale $\Lambda $ can be fixed requiring that the scalar fields are
starting to dominate the energy density of the universe today and that
both have already reached the tracking behavior.  The two conditions
are realized if
\begin{equation}
v(q_{0})\simeq \rho _{crit}^{0}\ ,\qquad \qquad v^{\prime \prime
}(q_{0})\simeq H_{0}^{2}\ , \label{conditions}
\end{equation}
where $\rho _{crit}^{0}=3M_P^{2}H_{0}^{2}/8\pi $ and $q_{0}$ are 
the present critical density and scalar fields VEV respectively. 
Eqs. (\ref{conditions}) imply
\begin{eqnarray}
\frac{\Lambda }{M_P} & \simeq &\left[ \frac{3}{4\pi
}\frac{(1+r)(3+r)}{(1-r)^{2}} \frac{1}{rN_c}\right] ^{\frac{1+r}{2(3-r)}}
\left( \frac{1}{2rN_{c}}\frac{\rho _{crit}^{0}}{M_P^{4}}\right)
^{\frac{1-r}{2(3-r)}}\ , \label{today} \\ && \nonumber \\
\frac{q_{0}^{2}}{M_P^{2}} & \simeq &\frac{3}{4\pi
}\frac{(1+r)(3+r)}{(1-r)^{2}} \frac{1}{rN_c}\ . \label{todayvev}
\end{eqnarray} 

Depending on the values for $N_{f}$ and $N_{c}$, $\Lambda $ and $
q_{0}/\Lambda $ assume widely different values. $\Lambda $ takes its
lowest possible values in the $N_{c}\rightarrow \infty $ ($N_{f}$
fixed) limit, where it equals \mbox{$4\cdot
10^{-2}(h^{2}/N_{f}^2)^{1/6}\ $GeV} (we have used $\rho
_{crit}^{0}/M_P^{4}=(2.5\cdot 10^{-31}h^{1/2})^{4}$).  For fixed
$N_{c}$, instead, $\Lambda $ increases as $N_{f}$ goes from $1$ to its
maximum allowed value, $N_{f}=1-N_{c}$.  For $N_{c}\gta 20$ and
$N_{f}$ close to $N_{c}$, the scale $\Lambda $ exceeds $M_P.$

The accuracy of the determination of $\Lambda $ given in (\ref{today})
depends on the present error on the measurements of $H_{0}$, {\it
i.e., } typically,{\it \ }$\delta \Lambda /\Lambda =\frac{1-r}{3-r}\ 
\delta H_{0}/H_{0}\lta 0.1$. 

In deriving the scalar potential (\ref{potscal}) and the field
equations (\ref{eom}) we have assumed that the system is in the
weakly coupled regime, so that the canonical form for the K\"{a}hler
potential may be considered as a good approximation. This condition is
satisfied as long as the fields' VEVs are much larger than the
non-perturbative scale $\Lambda $.  From eqs. (\ref{today}) and
(\ref{todayvev}), one can compute the ratio between the VEVs today and
$\Lambda $, and see that it is greater than unity for any $N_f$ as
long as $N_c \lta 20$. 

\subsection{Interaction with other cosmological fields}

The superfields $Q_{i}$ and $\overline{Q}_{i}$ have been taken as
singlets under the SM gauge group. Therefore, they may interact with
the visible sector only gravitationally, {\it i.e.  }via
non-renormalizable operators suppressed by inverse powers of the
Planck mass, of the form
\begin{equation}
\int d^{4}\theta \ K^{j}(\phi _{j}^{\dagger },\phi _{j})\ \cdot
\beta ^{ji}\left[ \frac{Q_{i}^{\dagger }Q_{i}}{M_P^{2}}
\right] \ \ , \label{coupling}
\end{equation}
where $\phi _{j}$ represents a generic standard model superfield. From
(\ref{todayvev}) we know that today the VEV's $q_{i}$ are typically
$O(M_P)$, so there is no reason to limit ourselves to the
contributions of lowest order in $|Q|^{2}/M_P^{2}$. Rather, we have
to consider the full (unknown) functions $\beta ^{ji} $
and the analogous $\overline{\beta }$'s for the $\overline{Q}_{i}$'s. 
Moreover, the requirement that the scalar fields are on the tracking
solution today, Eqs. (\ref{conditions}) implies that their mass is of
order $\sim H_{0}^{2}\sim 10^{-33}$ eV. 

The exchange of very light fields gives rise to long-range forces
which are constrained by tests on the equivalence principle, whereas
the time dependence of the VEV's induces a time variation of the SM
coupling constants\cite{car,dam}.  These kind of considerations sets
stringent bounds on the first derivatives of the $\beta ^{ji}$'s and
$\overline{\beta }^{ji}$'s {\it today,}

\[
\alpha ^{ji}\equiv \left.\frac{d\log \beta ^{ji}\left[ x_i^2 \right]
}{d x_i}\right|_{x_i=x_i^0} ,\qquad \qquad \overline{\alpha
}^{ji}\equiv \left.\frac{d\log \overline{\beta }^{ji}\left[ x_i^2
\right] }{d\,x_i}\right|_{x_i=x_i^0} \qquad,
\]
where $x_i \equiv q_i/M_P$.  To give an example, the best bound on the
time variation of the fine structure constant comes from the Oklo
natural reactor. It implies that $\left| \dot{\alpha}/\alpha \right|
<10^{-15}\ {\rm yr}^{-1}$ (see Ref.\cite{dam2}), leading to the following
constraint on the coupling with the kinetic terms of the
electromagnetic vector superfield $V$,
\begin{equation}
\alpha ^{Vi},\ \overline{\alpha }^{Vi}\ \lta\ 10^{-6}\ \frac{H_{0}}{
\left\langle \dot{q}_{i}\right\rangle }\,M_P \,, \label{decoupling}
\end{equation}
where $\left\langle \dot{q}_{i}\right\rangle $ is the average rate of
change of $q_{i}$ in the past $2\times 10^{9}{\rm yr}$. 

Similar --although generally less stringent-- bounds can be
analogously obtained for the coupling with the other standard model
superfields\cite{dam}. Therefore, in order to be phenomenologically
viable, any quintessence model should postulate that all the unknown
couplings $\beta ^{ji}$'s and $\overline{\beta }^{ji}$'s have a
common minimum close to the actual value of the $q_{i}$'s\footnote{
An alternative way to suppress long-range interactions, based on an
approximate global symmetry, was proposed in Ref.\cite{car}.}. 

The simplest way to realize this condition would be via the {\it least
coupling principle } introduced by Damour and Polyakov for the
massless superstring dilaton in Ref.\cite{dam3}, where a universal
coupling between the dilaton and the SM fields was postulated. In the
present context, we will invoke a similar principle, by postulating
that $\beta ^{ji}=\beta $ and $\overline{\beta }^{ji}=\overline{\beta
}$ for any SM field $\phi _{j}$ and any flavor $i$. For simplicity, we
will further assume $\beta =\overline{\beta }$ . 

The decoupling from the visible sector implied by bounds like (\ref
{decoupling}) does not necessarily mean that the interactions between
the quintessence sector and the visible one have always been
phenomenologically irrelevant. Indeed, during radiation domination the
VEVs $q_{i}$ were typically $\ll M_P$ and then very far from the
postulated minimum of the $\beta $'s. 
This leads, for the quintessence fields, to the generation of SUSY breaking 
masses (proportional to $H$) by the same mechanism discussed by Dine, Randall, 
and Thomas in Ref.\cite{drt}.

The main phenomenological effect of these time-dependent SUSY breaking 
masses is to prevent the fields from taking too large values. This 
results in an improved attraction to the tracker solution (for further 
details on this point see Ref.\cite{noi}).

\subsection{Numerical Results}

The general results of the previous discussion are illustrated in
the figures for the particular case $N_f=2$, $N_c=6$.

\begin{figure}
\begin{center}
\vspace*{13pt}
\epsfig{figure=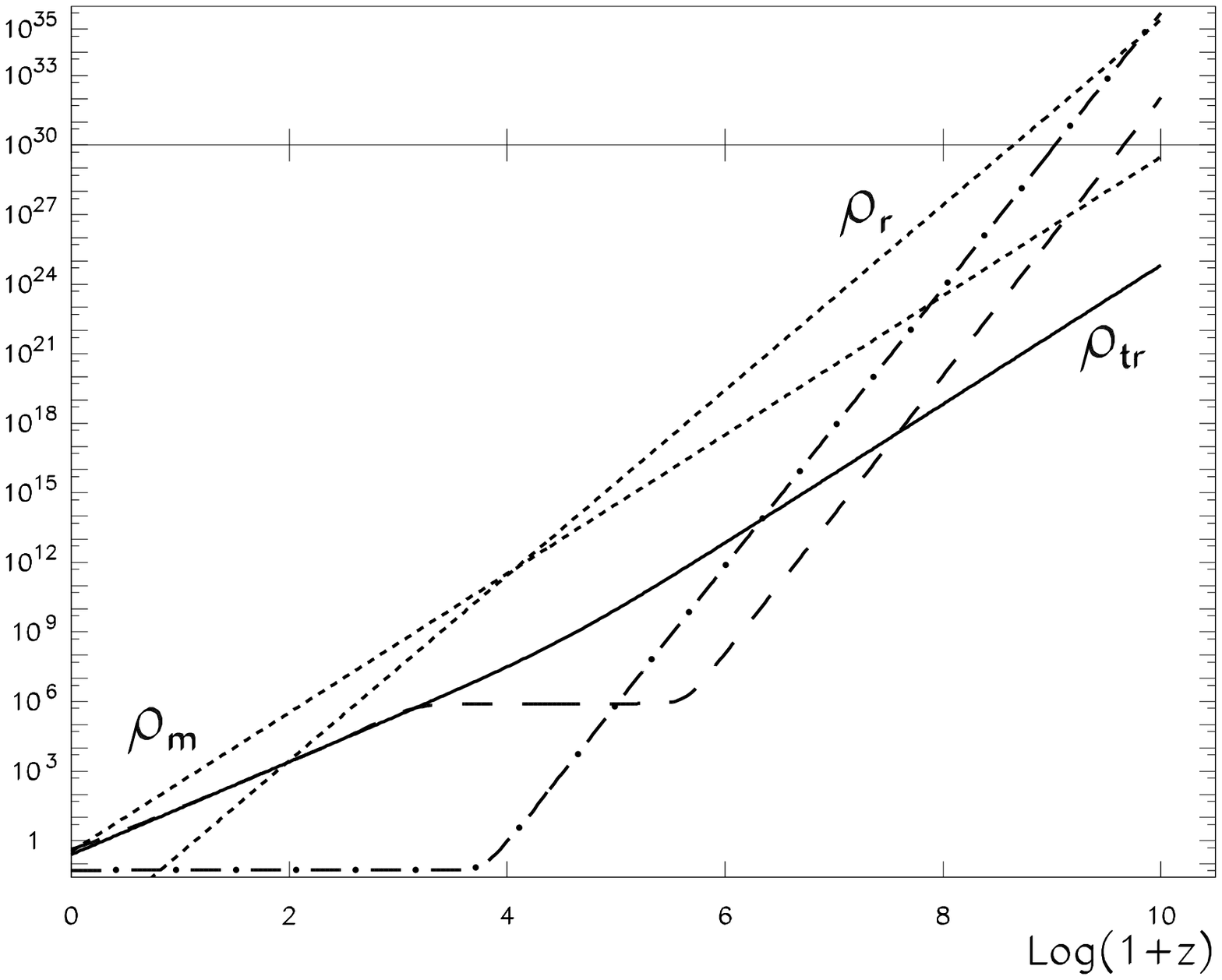,bbllx=30,bblly=200,bburx=560,bbury=600,width=9.0cm}
\vspace*{0.2truein}		
\fcaption{The evolution of
the energy densities $\rho$ of different cosmological components is given 
as a funcion of red-shift. All the energy densities are normalized to the 
present critical energy density $\rho_{cr}^0$. Radiation and matter
energy densities are represented by the short-dashed lines, whereas the
solid line is the energy density of the tracker solution discussed in
Section 2.3. The long-dashed line is the evolution of the scalar field
energy density for a solution that reaches the tracker before the present 
epoch; while the dash-dotted line represents the evolution for a solution 
that overshoots the tracker to such an extent that it has not yet had enough 
time to re-join the attractor.}
\end{center}
\end{figure}

In Fig.1 the energy densities {\em vs.}  redshift are given. We have
chosen the same initial conditions for the two VEVs, in order to
effectively reproduce the one-scalar case of eq. (\ref{onescalar}),
already studied in Refs.\cite{rp,liddle,swz}. 
No interaction with other fields of the type discussed in the
previous section has been considered. 

We see that, depending on the initial energy density of the scalar
fields, the tracker solution may (long dashed line) or may not 
(dash-dotted line) be reached before the present epoch. The latter case
corresponds to the overshoot solutions discussed in Ref.\cite{swz},
in which the initial scalar field energy is larger than $\rho_B$ and 
the fields are rapidly pushed to very large values. 
The undershoot region, in which the energy density is always
lower than the tracker one, corresponds to 
$\rho_{cr}^0 \leq \rho_Q^{in} \leq \rho_{tr}^{in}$. 
Thus, all together, there are around 35  orders of magnitude in $\rho_Q^{in}$ 
at redshift $z+1 = 10^{10}$ for which the tracker solution is reached 
before today. Cleary, the more we go backwards in time, the larger is the
allowed initial conditions range. 

Next, we explore to which extent the two-field system that we are
considering behaves as a one scalar model with inverse power-law
potential. In Fig. 2 we plot solutions with the same initial energy
density but different ratios between the initial values of the two
scalar fields. 
Given any initial energy density such that -- for $q^{in}_1/q^{in}_2 =1$ -- the
tracker is joined before today, there always exists a limiting value for the
fields' difference above which the attractor is not reached in time.

\begin{figure}
\begin{center}
\vspace*{13pt}
\epsfig{figure=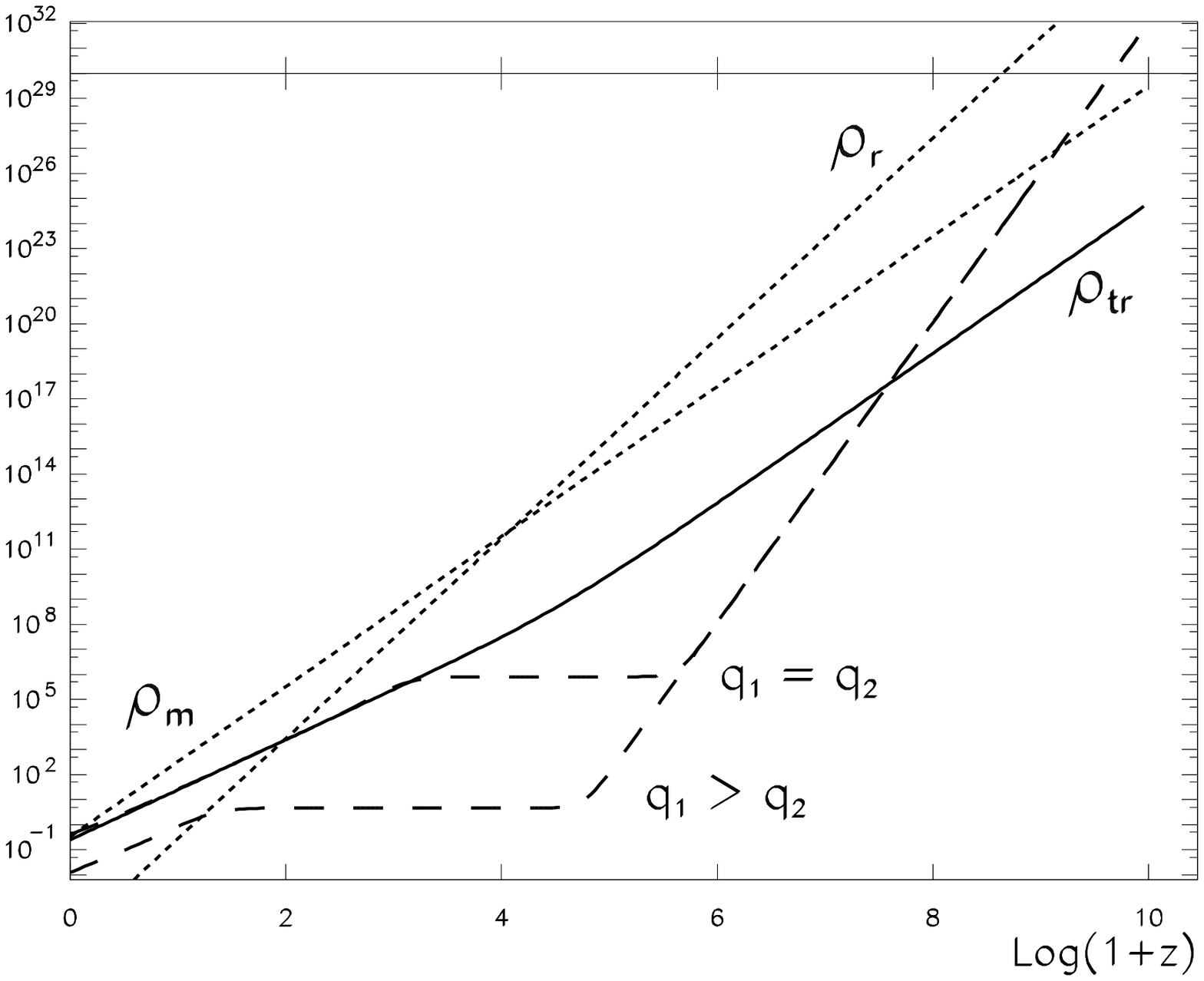,bbllx=30,bblly=200,bburx=560,bbury=600,width=9.0cm}
\vspace*{0.2truein}		
\fcaption{The effect of taking different initial conditions
for the fields, at the same initial total field energy.
Starting with $q_1^{in}/q_2^{in} =10^{15}$ 
the tracker behaviour is not reached today. For comparison we plot the
solution for $q_1^{in}/q_2^{in} =1$.}
\end{center}
\end{figure}

\section{Aknowledgments}

We thank Massimo Pietroni with whom we obtained most of the results reported 
in this talk. We are also grateful to the organizer of this Workshop, 
Prof. Milla Baldo Ceolin, for the nice and stimulating atmosphere in which
the meeting took place.

\end{document}